\documentclass[prd,reprint,nofootinbib,showpacs,superscriptaddress]{revtex4-1}
\usepackage{graphicx} 
\usepackage{hyperref}
\usepackage{amsfonts}
\usepackage{amsmath,amssymb}
\usepackage{bm} 
\usepackage{color}
\usepackage{epstopdf} 
\usepackage{epsfig}
\usepackage{subfig} 

{\rm }

\def\be{\begin{equation}}
 \def\ee{\end{equation}}
 \def\bea{\begin{eqnarray}}
 \def\eea{\end{eqnarray}}
 \def\bes{\begin{eqnarray}}
 \def\ees{\end{eqnarray}}
 \def\bi{\begin{itemize}}
 \def\ei{\end{itemize}} 

\def\2{\frac{1}{2}}
\def\4{\frac{1}{4}}

\begin{document}

\title{Loss-tolerant position-based quantum cryptography
}

\author{Bing Qi}
\email{qib1@ornl.gov}
\affiliation{Quantum Information Science Group, Computational Sciences and Engineering Division,
Oak Ridge National Laboratory, Oak Ridge, Tennessee 37831-6418, USA}
\affiliation{
 Department of Physics and Astronomy, The
University of Tennessee, Knoxville, Tennessee 37996-1200, USA
}
\author{George Siopsis}
\email{siopsis@tennessee.edu}

\affiliation{
 Department of Physics and Astronomy, The
University of Tennessee, Knoxville, Tennessee 37996-1200, USA
}

\date{\today}

\begin{abstract}

Position-based quantum cryptography (PBQC) allows a party to use its geographical location as its only credential to implement various cryptographic protocols. Such a protocol may lead to important applications in practice. Although it has been shown that any PBQC protocol is breakable if the adversaries pre-share an arbitrarily large entangled state, the security of PBQC in the bounded-quantum-storage model is still an open question. In this paper, we study the performance of various PBQC protocols over a lossy channel under the assumption that no entanglement is pre-shared between adversaries. By introducing the decoy state idea, we show that an extended BB84-type PBQC protocol implemented with a weak coherent source and realistic single photon detectors can tolerate an overall loss (including both the channel loss and the detection efficiency) of $13$ dB if the intrinsic quantum bit error rate is $1$\%. We also study a few continuous variable PBQC protocols and show that they suffer from a 3 dB loss limitation.
\end{abstract}

\maketitle

\section{Introduction}

Quantum physics has been applied to enhance communication security. A well-known example is quantum key distribution (QKD) \cite{BB84,E91,Gisin02,Scarani09,Lo14}, which has been proved to be unconditionally secure in principle \cite{Mayers01,Lo99,Shor00}. More recently, position-based quantum cryptography (PBQC) has drawn a lot of attention \cite{Kent06, Kent11, Kent112, Malaney10, Lau11, Buhrman14, Tomamichel09}.

In PBQC, a legitimate party uses its geographical location as its only credential to implement various cryptographic protocols. One fundamental task in PBQC is secure position verification, in which an honest prover tries to prove to a set of trusted verifiers that he/she is at a certain location. It has been shown that, given a secure position verification scheme, other position-based cryptographic tasks, such as authentication and key distribution, can also be implemented \cite{Buhrman14}.

Although previous studies have shown that any PBQC protocols are breakable if the adversaries pre-share an arbitrarily large entangled state \cite{Buhrman14}, the security of PBQC in the bounded-quantum-storage model, which is more relevant in practice, is still an open question. In \cite{Buhrman14}, the security of PBQC has been proved based on the assumption that no entanglement is pre-shared between adversaries. In \cite{Tomamichel09}, the authors proposed a one-round quantum position verification scheme and proved its security against adversaries with an amount of pre-shared entanglement that is linear in the number of qubits transmitted. In contrast, all classical position-based cryptographic protocols are breakable without using pre-shared entanglement \cite{Chandran09}. This highlights the advantage of PBQC over its classical counterpart.

One important issue in PBQC which has not been carefully addressed is its tolerance to the loss of the quantum communication channel. In most of the previous studies, the BB84 encoding scheme has been adopted to implement a PBQC protocol. It is easy to show (see Section \ref{sec:2}) that the above protocol is insecure as long as the overall loss is above $3$ dB. In practice, the loss of the measurement device itself (including optical coupling loss and detector efficiency) is typically above $3$ dB. This suggests the BB84-type PBQC protocols are impractical.

In this paper, we explore PBQC protocols which can go beyond the above 3 dB limit. We study the performance of various PBQC protocols over a lossy channel under the assumption that no entanglement is pre-shared between adversaries. These PBQC protocls are derived naturally from the corresponding QKD schemes, including both discrete-variable (DV) and continuous-variable (CV) schemes. Our study shows that a DV PBQC protocol, which was proposed in \cite{Kent11, Lau11}, can tolerate an arbitrarily high loss in the noiseless case. By introducing the decoy state idea \cite{Decoy}, the above protocol can tolerate an overall loss of 13 dB at a quantum bit error rate (QBER) of $1$\%, even when implemented with a practical weak coherent source and realistic single photon detectors (SPDs). In contrast, all the CV PBQC protocols studied in this paper suffer from a $3$ dB loss limitation, even by using arbitrarily squeezed states and noiseless detectors.

This paper is organized as follows: in Section \ref{sec:2}, we outline the basic procedures of a PBQC protocol and the main assumptions we adopt.  In Section \ref{sec:DV}, we study the performance of a DV PBQC protocol in both ideal and realistic scenarios. In Section \ref{sec:CV}, we present our results of CV PBQC protocols.  We conclude with a brief discussion in Section \ref{sec:dis}.

\section{Position-based quantum cryptography}
\label{sec:2}

For simplicity, in this paper, we only consider the one-dimensional case with two verifiers, $V_0$ and $V_1$, and a prover $P$ in the middle. The verifiers wish to verify the position of $P$. Unfortunately, there are two adversaries, $E_0$ (between $V_0$ and $P$) and $E_1$ (between $V_1$ and $P$) who will try to fake $P$. 

We assume that $V_0$ and $V_1$ hold synchronized clocks and they also share a secure communication channel (which could be established by using QKD). $E_0$ and $E_1$ have full control of the communication channels (both quantum and classical) between $V_0$ and $P$, as well as the one between $V_1$ and $P$. Furthermore, they have perfect detectors (noiseless and lossless) and can perform any local operations allowed by quantum mechanics with arbitrarily high precision. However, $E_0$ and $E_1$ do not share any entanglement. Note that in the above picture, the adversaries can completely hijack signals from $P$. So, we can simply ignore the existence of $P$ in our discussion below.

Most existing PBQC protocols are based on the BB84-type coding scheme. The basic procedure is summarized below:
\begin{enumerate}
\item The verifiers agree on random bits $x, \theta \in \{ 0,1 \}$. $V_0$ prepares a qubit in the state
\be |\psi \rangle = H^\theta |x\rangle \ee
where $|0\rangle$ and $|1\rangle$ are computational basis states, and $H$ is the Hadamard matrix.
\item $V_0$ sends $|\psi\rangle$ to $P$ through a free-space quantum channel, and $V_1$ sends $\theta$ to $P$ through a free-space classical channel. $V_0$ and $V_1$ coordinate their transmission time to make sure that $|\psi\rangle$ and $\theta$ arrive at $P$ at the same time.
\item As soon as $|\psi\rangle$ and $\theta$ arrive, $P$ performs a measurement in the basis $\{ H^\theta |0\rangle , H^\theta |1\rangle \}$ and broadcasts the measurement result $x'$ back to $V_0$ and $V_1$ immediately. 
\item If the verifiers receive $x'$ at the time consistent with the position of $P$, \emph{and} $x'$ agrees with $x$, the location of $P$ is accepted as authenticated.
\end{enumerate}
Intuitively, the security of the above protocol can be understood from the quantum no-cloning theorem \cite{No-cloning}: after step 2, $E_0$ can intercept $|\psi\rangle$ and $E_1$ can intercept $\theta$ before $P$ receives them. However, without the other piece of information, neither of them can perform step 3. While $E_1$ can keep one copy of the classical information $\theta$ and forward another copy to $E_0$, $E_0$ cannot make perfect copies of the quantum state $|\psi\rangle$. If $E_0$ keeps $|\psi\rangle$ in a quantum memory and performs the measurement after she receives $\theta$ from $E_1$, it will be too late to report the measurement result back to $V_1$. On the other hand, if $E_0$ forwards $|\psi\rangle$ to $E_1$, who in turn performs the measurement, it will be too late to report the measurement result back to $V_0$. We remark that the above argument is fallacious if $E_0$ and $E_1$ pre-share entanglement, which allows them to perform quantum teleportation \cite{Bennett93}. In this paper, we will only consider the case of no pre-shared entanglement between adversaries.

One major issue in practice is the low reporting rate expected from honest $P$. Due to the channel loss and the low efficiency of practical SPDs, most of the time, $P$ will report a ``no detection'' event. The adversaries can take advantage of the low detection probability and fool the verifiers. More specifically, once $E_0$ receives $|\psi\rangle$, she randomly chooses one of the two bases and performs a measurement. Then she forwards a copy of her basis information and the measurement result to $E_1$. On the other hand, once $E_1$ receives $\theta$, she forwards a copy to $E_0$. If $E_0$ happens to use the correct basis, $E_0$ and $E_1$ report the measurement result to $V_0$ and $V_1$, respectively. Otherwise, they claim no detection. Since there are only two possible bases in the above protocol, the adversaries can achieve a $50\%$ reporting rate without introducing errors. This suggests that the above protocol is insecure as long as the overall loss is above $3$ dB \footnote{The verifiers cannot distinguish between the channel loss and the detection loss in a PBQC protocol. So we use only one parameter, the overall loss, to quantify the transmittance. In this paper, we also neglect imperfection in $V_0$'s source, which cannot be accessed by the adversaries, and can be well calibrated by the verifiers.}. Considering the high coupling loss of a free-space communication channel and the low detection efficiency of a practical SPD, the above PBQC protocol may only work over a very short distance. To make PBQC useful in practice, we need protocols which can go beyond this $3$ dB limit. In the rest of this paper, we study the performance of various PBQC protocols over a lossy channel under the assumption that no entanglement is pre-shared between adversaries.

\section{Discrete-variable PBQC protocol}
\label{sec:DV}

One approach to the development of a loss-tolerant PBQC protocol is to employ more than two encoding bases. Here we study the performance of a PBQC protocol based on a natural extension of the BB84 coding scheme, where $V_0$ employs $K$ encoding bases uniformly distributed over the whole Bloch sphere. Compared to the original PBQC scheme, this protocol also has an enhanced security \cite{Lau11}.

We first describe the procedure of the PBQC protocol. For simplicity, we assume that $K$ is large enough that the encoding bases can be treated as uniformly and  continuously distributed over the Bloch sphere. 

\begin{enumerate}
\item The verifiers agree on a random bit $x \in \{ 0,1 \}$ and two continuous random numbers $\theta\in [0,\pi]$, $\phi\in [0,2\pi)$.

$V_0$ prepares a qubit in the state
\be |\psi \rangle = U(\theta,\phi) |x\rangle \ee
where
$ U(\theta,\phi) = \left( \begin{array}{cc} \cos\frac{\theta}{2} & e^{i\phi}\sin\frac{\theta}{2} \\
 -e^{-i\phi}\sin\frac{\theta}{2} & \cos\frac{\theta}{2} \end{array} \right) $
\item $V_0$ sends $|\psi\rangle$ to $P$, and $V_1$ sends $\theta$ and $\phi$ to $P$. $V_0$ and $V_1$ coordinate their transmission times to make sure that $|\psi\rangle$, $\theta$ and $\phi$ arrive at $P$ at the same time.
\item As soon as $|\psi\rangle$, $\theta$ and $\phi$ arrive, $P$ performs a measurement in the basis $\{ U(\theta,\phi) |0\rangle , U(\theta,\phi) |1\rangle \}$. If $P$ detects a photon successfully, he broadcasts the measurement result $x'$ back to $V_0$ and $V_1$ immediately. Otherwise, he reports no detection.
\item If the verifiers receive $x'$ at a time inconsistent with the position of $P$, the protocol fails and will be terminated.
\item Through an authenticated classical channel, $V_0$ and $V_1$ compare the reported measurement results received by them. If they receive different results, the protocol fails and will be terminated.
\item $V_0$, $V_1$ and $P$ repeat the above procedures many times. They estimate the reporting rate and the QBER. For a given reporting rate, if the QBER is below certain predetermined value, the location of $P$ is accepted as authenticated.
\end{enumerate}

As noted in \cite{Lau11}, noisy operation in the honest case do not produce inconsistent results between different verifiers. This is because the classical communication channel used by $P$ to broadcast the measurement results is virtually noiseless. By performing step 5, we can further limit the adversaries' power. For example, one possible hacking strategy is as follows. As soon as $E_1$ receives $\theta$ and $\phi$, she forwards a copy to $E_0$. As soon as $E_0$ receives $|\psi\rangle$, she applies an optimal cloning machine to generate two imperfect copies of the input quantum state, keeps one and sends the other one to $E_1$. Once $E_0$ receives the basis information from $E_1$, she measures her copy and reports the measurement result to $V_0$. Similarly, $E_1$ measures her own copy and reports the measurement result to $V_1$. In general, such an attack will produce inconsistent results between $V_0$ and $V_1$ and thus can be detected in step 5. The same argument can also be applied in CV PBQC protocols based on coherent detection.

Below, we will study the performance of the above protocol implemented with either a perfect single photon source or a more practical weak coherent source.

\subsection{Single photon source}

We assume that $V_0$ employs a perfect single photon source to implement the above PBQC protocol. If the PBQC system is noiseless, the verifiers expect a perfect correlation between the measurement result reported by the honest $P$ and the information encoded by $V_0$. However, if the overall transmittance of the quantum communication system between $V_0$ and $P$ is $\eta$, then the verifiers expect a reporting rate of $\eta$.

Having no information about the basis, $E_0$ will make a choice of basis $(\theta_E, \phi_E)$ for her measurement. 
To make the attack undetectable, $E_0$ can randomly choose one of the above $K$ bases. $E_0$ will then measure the quantum state sent by $V_0$, and forward a copy of her basis information and the measurement result to $E_1$. If $E_0$ happens to use the correct basis, $E_0$ and $E_1$ will report the measurement result to $V_0$ and $V_1$ correspondingly. Otherwise, they claim no detection. Obviously, the reporting rate $R$ is given by $2/K$ (the adversaries report the result of the measurement, if $\theta = \phi =0$, and the opposite result, if $\theta = \pi$, $\phi =0$). If $R<\eta$, Eve's attack can be detected due to the abnormal reporting rate. As $K\rightarrow\infty$, the adversaries' reporting rate $R\rightarrow0$, which suggests that the above protocol can tolerate arbitrarily high channel loss in the noiseless case.

In practice, the quantum communication system between $V_0$ and $P$ is noisy due to imperfect state preparation, detector noise, etc. In this case, the expected QBER from the honest $P$ will not be zero. The adversaries can take advantage of the non-zero QBER to further improve their attack. As long as the adversaries can produce the same reporting rate and the QBER expected from the honest $P$ at a time consistent with the position of $P$, the security of the PBQC protocol is compromised.

The hacking strategy is discussed in detail in the Appendix in the case of a finite $K$. Here, we proceed with a description of the strategy in the large $K$ limit.

As soon as $E_0$ receives the quantum state $|\psi\rangle$, she measures it in a basis randomly chosen from the whole Bloch sphere\footnote{If $E_0$ always chooses the same basis, the reporting probability will depend on $V_0$'s basis. This can be detected by the verifiers.} and forwards a copy of her basis information $(\theta_E,\phi_E)$ and the measurement result ($x'$) to $E_1$. Since the $K$ bases chosen by the verifiers are uniformly distributed, for simplicity, in our calculations we set $\theta_E = \phi_E =0$, without loss of generality. The error probability of $E_0$'s measurement result is given by $\sin^2 \frac{\theta}{2}$, which is independent of $\phi$. The adversaries' strategy is as follows: if $\theta\in [0,\Theta_0)$ (where $\Theta_0<\pi/2$ is a constant predetermined by the adversaries), they will report $x'$ to $V_0$ ($V_1$); if $\theta\in (\pi-\Theta_0, \pi]$, they will flip the bit and report $x'+1$ (mod 2) to $V_0$ ($V_1$); otherwise, they claim no detection. 

It is easy to show that the adversaries' reporting rate is given by
\be\label{eq3} R_1 = \int_0^{\Theta_0} \sin\theta d\theta =2\sin^2 \frac{\Theta_0}{2} \ee 
and the average QBER introduced by the above attack is given by
\be\label{eq4} Q_1 = \frac{1}{R_1} \int_0^{\Theta_0} \sin^2 \frac{\theta}{2} \sin\theta d\theta = \frac{R_1}{4} \ee
In the above equations, the subscript $1$ indicates that a single photon source is applied.
 
Eq.\ \eqref{eq4} quantifies the relation between the maximum reporting rate $R_1$ and the corresponding QBER $Q_1$ due to the attack. It also shows how much loss the PBQC protocol can tolerate given an intrinsic QBER. For example, if the intrinsic QBER of the legitimate quantum communication system is $1$\%, then the adversaries could reproduce the same QBER with a reporting rate of up to $4$\%. This reporting rate is consistent with the one expected from a quantum channel with an overall loss of $14$ dB. This implies that given a $1$\% intrinsic QBER, this protocol is secure as long as the overall loss is below $14$ dB. In general, the lower the intrinsic QBER, the higher the tolerable loss.

In the Appendix, we study the case of a finite number $K$ of uniformly distributed bases and obtain the maximum reporting rate for a given QBER. As we show there, the above results are recovered in the large $K$ limit. The convergence is actually fast, indicating that for practical applications $K$ need not be very large.

\subsection{Weak coherent source}

A high quality, efficient single photon source is not available in practice. In this subsection, we will study the performance of the above PBQC protocol implemented with a practical attenuated laser source. We assume that the global phase of each laser pulse has been randomized properly, so we can model the source as a phase-randomized weak coherent state. 

The performance of the BB84 QKD protocol degrades significantly when implemented with a phase-randomized weak coherent state. This is mainly due to the photon number splitting (PNS) attack \cite{PNS}: if the quantum signal contains more than one photons, an eavesdropper can split out one photon, store it in a quantum memory, and perform a measurement when the basis information is available. One way to solve the above problem is to introduce the decoy state idea \cite{Decoy}. As we will show below, the decoy state idea can also be applied in PBQC protocols.

The photon number of a coherent state follows a Poisson distribution:
\be P_n=\dfrac{\mu^{n}_1}{n!}e^{-\mu_1} \ee
where $\mu_1$ is the average photon number of the coherent state, which is determined by $V_0$.

We assume that the quantum communication system between $V_0$ and $P$ is characterized by the following parameters: $\eta$ -- the overall transmittance; $Y_0$ -- the dark-count probability of SPD; and $e_{\text{det}}$ -- the error probability due to misalignment.

In normal conditions (no adversaries), the expected reporting rate and QBER from the honest $P$ are given, respectively, by \cite{Ma05}
\bes\label{eq12} R^{(\mu_1)}_P &=& Y_0+1-e^{-\eta\mu_1} \nonumber\\
Q^{(\mu_1)}_P &=& \frac{\dfrac{1}{2}Y_0+e_{\text{det}}(1-e^{-\eta\mu_1})}{R^{(\mu_1)}_P} \ees
where we have assumed that the error rate of the dark count is 0.5.

The adversaries' hacking strategy is as follows. As soon as $E_0$ receives the quantum state from $V_0$, she performs a quantum non-demolition measurement to determine the photon number $n$ in $V_0$'s signal. Then there are three possibilities:
\bi\item[{\em (a)}] If $n=0$, $E_0$ and $E_1$ report no detection.
\item[{\em (b)}] If $n=1$, $E_0$ and $E_1$ perform the same attack as described in the previous subsection.
\item[{\em (c)}] If $n>1$, $E_0$ can optimize her measurement strategy to gain maximum information about $x$. \footnote{In the case of $n>1$, one may suggest the following attack: $E_0$ keeps one photon in her quantum memory and forwards the rest to $E_1$. After receiving the basis information from $E_1$, $E_0$ can measure her photon in the correct basis and report the measurement result to $V_0$. Similarly, $E_1$ can measure the photons forwarded by $E_0$ and report the measurement result to $V_1$. In the ideal case, the adversaries will not introduce errors. However, in practice, the quantum state prepared by $V_0$ cannot be perfect (or $V_0$ may intentionally introduce a small amount of noise). So the above attack will produce inconsistent results received by $V_0$ and $V_1$, and thus can be detected in step 5 of the protocol.}
\ei
We define $R_n$ and $Q_n$ as the conditional reporting rate and QBER, respectively, given that $V_0$'s signal contains $n$ photons. In the attack outlined above, $R_0=Q_0=0$, whereas for a given $Q_1$, $R_1$ is upper bounded by Eq.\ \eqref{eq4}. In the case of $n>1$, determining the optimal measurement strategy for $E_0$ is a problem related to the problem of optimal state estimation with post-measurement information \cite{Ballester08}. The latter is by itself an interesting research topic. Here we will simply assume that $R_n=1$ and $Q_n=0$ for $n>1$, which is the most favorable assumption for the adversaries, although a corresponding strategy may not exist. The overall reporting rate and QBER under the above assumptions are given by, respectively,
\bes\label{eq13} R^{(\mu_1)} &=& \sum_{n=0}^\infty P_n R_n=\mu_{1} e^{-\mu_1} R_1+1-(1+\mu_{1}) e^{-\mu_1}\nonumber\\
Q^{(\mu_1)} &=& \dfrac{\sum_{n=0}^\infty P_n R_n Q_n}{\sum_{n=0}^\infty P_n R_n}= \dfrac{\mu_{1} e^{-\mu_1} {R_1}^2}{4R^{(\mu_1)}}\ees
where we used the relation $Q_1=R_1/4$ (Eq.\ \eqref{eq4}).

Without using decoy states, $R^{(\mu_1)}$ and $Q^{(\mu_1)}$ are the only parameters available to the verifiers. To make their attack undetectable, the adversaries need to match $R^{(\mu_1)}$ and $Q^{(\mu_1)}$ with $R^{(\mu_1)}_P$ and $P^{(\mu_1)}_P$ by adjusting $R_1$. We performed a numerical simulation using Eqs.\ \eqref{eq12} and \eqref{eq13} with the following system parameters: $Y_0=10^{-5}$, $e_{\text{det}}=0.01$, and $\mu_1=0.018$ ($\mu_1$ has been optimized to maximize the tolerable overall loss). Our simulation shows that as long as the overall transmittance $\eta<0.07$ (which corresponds to $11.5$ dB loss), the above PBQC protocol is insecure. This shows that the performance of the PBQC implemented with a weak coherence source is worse than the one implemented with a perfect single photon source.

The basic idea of the decoy-state protocol is quite simple. For each transmission, $V_0$ randomly chooses a number from a predetermined set $\{\mu_1,\mu_2,\dots,\mu_M\}$, and sets the average photon number of the weak coherent pulse to the corresponding value. At the end of the protocol, $V_0$ can determine the reporting rate and QBER for each $\mu_i$ $(i=1,2...M)$ separately. Now the parameters available to the verifiers are in the set $\{ R^{(\mu_1)}, Q^{(\mu_1)}; R^{(\mu_2)}, Q^{(\mu_2)};\dots ; R^{(\mu_M)}, Q^{(\mu_M)} \}$. As $M\rightarrow\infty$, the verifiers can determine the reporting rate and QBER of the single-photon pulse, $(R_1, Q_1)$, precisely \cite{Decoy}. This is equivalent to the case that $V_0$ has a perfect single photon source. So, by introducing the decoy-state idea, in the asymptotic case (i.e., $M\rightarrow\infty$, and the finite-date-size effect can be ignored) the performance of PBQC implemented with a weak coherence source approaches to the one implemented with a perfect single-photon source.

In practice, it is inefficient to use a large $M$. Here, we study the simplest case where $M=2$. Similar to \eqref{eq12}, the expected reporting rate and QBER when $V_0$ chooses $\mu_2$ as the average photon number are given, respectively, by  
\bes\label{eq14} R^{(\mu_2)}_P&=& Y_0+1-e^{-\eta\mu_2} \nonumber\\
Q^{(\mu_2)}_P &=& \frac{\dfrac{1}{2}Y_0+e_{\text{det}}(1-e^{-\eta\mu_2})}{R^{(\mu_2)}_P} \ees
Using \eqref{eq12} and \eqref{eq14}, we can determine a lower bound of the reporting rate of a single-photon pulse \cite{Ma05}:
\begin{multline}\label{eq15}
R^{(1)}_L=\dfrac{\mu_1}{\mu_1\mu_2-\mu^{2}_2}\left( R^{(\mu_2)}_{P} e^{\mu_2}-R^{(\mu_1)}_{P} e^{\mu_1}\dfrac{\mu^{2}_2}{\mu^{2}_1} \right. \\ \left. - 2Q^{(\mu_1)}_{P} R^{(\mu_1)}_{P} e^{\mu_1}\dfrac{\mu^{2}_1-\mu^{2}_2}{\mu^{2}_1} \right) 
\end{multline}
By using $R^{(1)}_L$ as an estimation of $R_1$ in \eqref{eq13}, we can determine a lower bound of the QBER due to the attack:
\be\label{eq16} Q^{(\mu_1)}_L=\dfrac{\mu_{1} e^{-\mu_1} (R^{(1)}_L)^{2}}{4R^{(\mu_1)}}\ee
To make their attack undetectable, the adversaries need to match $R^{(\mu_1)}$ and $Q^{(\mu_1)}_L$ with $R^{(\mu_1)}_P$ and $Q^{(\mu_1)}_P$, respectively. We performed a numerical simulation using Eqs.\ \eqref{eq12}, \eqref{eq13}, \eqref{eq14}, \eqref{eq15}, and \eqref{eq16} with the following system parameters: $Y_0=10^{-5}$, $e_{\text{det}}=0.01$, $\mu_1=0.12$, and $\mu_2=0.1$. Our simulation shows that as long as the overall transmittance $\eta>0.05$ (which corresponds to about 13 dB loss), the above PBQC protocol is secure. This result is comparable with the one based on a perfect single photon source. Furthermore, compared with the weak coherent state protocol without decoy states, the decoy state protocol allows $V_0$ to use a relatively large average photon number, which is more efficient.

\section{Continuous-variable PBQC protocols}
\label{sec:CV}

Continuous-variable (CV) QKD protocols based on optical coherent detection (such as homodyne or heterodyne detection) have been demonstrated as useful solutions for secure key distribution \cite{Hillery00,Grosshans03,Lodewyck05,Qi07,Jouguet13}. While the SPD used in DV QKD may output a ``no-detection'' event, the homodyne (or heterodyne) detector used in CV QKD always yields a measurement result regardless of channel loss. This suggests the ``post-selection'' strategy discussed in the previous Section, where the adversaries report a measurement result only when they happen to use a ``good'' measurement basis,  cannot be applied in CV PBQC protocols based on coherent detection. Another distinct advantage of the CV protocol is that the local oscillator (LO) employed in coherent detection acts as a ``mode selector'' which can suppress broadband noise photons effectively \cite{Qi10}. This feature is especially appealing in free-space quantum communication systems where the background noise due to ambient light is high.

One of the most successful CV QKD protocols is the Gaussian-modulated-coherent-state (GMCS) QKD \cite{Grosshans03}. In this protocol, Alice draws two random numbers, $x_A$ and $p_A$, from a set of Gaussian random numbers, and sends a coherent state $|x_A+ip_A\rangle$ to Bob. Bob randomly chooses to measure either the amplitude quadrature ($q_X$) or the phase quadrature ($q_P$) by performing an optical homodyne detection. Later on, Bob informs Alice of which quadrature he measured for each transmission, and then they can generate a secure key from the corresponding data.
 
The above GMCS QKD protocol can be extended into a PBQC protocol as follows. $V_0$ prepares and sends the coherent state $|x_A+ip_A\rangle$ to $P$, while $V_1$ sends $P$ a random number $\theta$ which is uniformly distributed in $[0,\pi]$. Once $P$ receives the quantum state from $V_0$, he performs a homodyne detection along the phase $\theta$, and reports back the measurement result of the quadrature
\be\label{eq17} q_{\theta}=q_X \cos \theta  + q_P \sin \theta \ee
If the verifiers receive the measurement results at a time consistent with the position of $P$, and the measurement noise is consistent with the one expected from the channel with a predetermined loss, the location of $P$ is accepted as authenticated.

Unfortunately, it is easy to show that the above PBQC protocol is insecure as long as the overall loss is above $3$ dB. The hacking strategy is as follows. As soon as $E_0$ receives the quantum state from $V_0$, she splits it into two parts ($I$ and $II$) using a 50:50 beam splitter. She then measures the amplitude quadrature of part $I$ and  the phase quadrature of part $II$, and forwards the measurement results, $(q_X, q_P)$, to $E_1$. Once $E_0$ receives the information of $\theta$ forwarded by $E_1$, she calculates $q_{\theta}$ (Eq.\ \eqref{eq17}), and reports it  to $V_0$. $E_1$ performs the same calculation and reports the result to $V_1$. It is easy to verify that the noise due to this attack is equivalent to the one due to a channel with $3$ dB loss. So the above attack will always succeed, as long as the overall loss is above $3$ dB.

Can we improve the performance of PBQC by using squeezed states? Suppose that $V_0$ prepares a squeezed state with a randomly chosen squeezing angle $\theta$. By using an infinitely squeezed state, a quadrature measurement in any angle other than $\theta$ will result in an arbitrarily large uncertainty. If $E_0$ measures both the amplitude quadrature $q_X$ and the phase quadrature $q_P$ as described above, most likely, the uncertainties in both measurements are very high. Intuitively, this will result in a large uncertainty in $E_0$'s estimation of $q_{\theta}$ (Eq.\ \eqref{eq17}). It follows that the adversaries will not be able to reproduce the small measurement noise expected from the honest $P$, and thus they will be caught by the verifiers. However, as we will show below, the above intuition is wrong. The performance of the above coherent-state PBQC protocol is not improved by using squeezed states. Roughly speaking, this is due to the fact that the noises in $q_X$ and $q_P$ are correlated and can be canceled out when $E_0$ estimates $q_{\theta}$.

The squeezed state PBQC protocol is summarized as follows.

\begin{enumerate}
\item The verifiers agree on two random numbers $\theta$ and $\alpha$, with $\theta$ uniformly distributed in $[ 0, \pi]$, and $\alpha$ drawn from a Gaussian distribution of mean $0$ and variance $\sigma^2$.
$V_0$ prepares the squeezed coherent state
\be |\psi\rangle = R(\theta)  D(\alpha) S (s) |0\rangle \ee
where $R(\theta)=e^{i\theta a^\dagger a}$ represents a rotation, $D(\alpha) = e^{\alpha (a^\dagger - a)}$ represents a displacement, and $S(s) = e^{\frac{s}{2} (  a^2- a^{\dagger2})}$ is the squeezing operator.
We assume that $V_0$ can generate a squeezed state with an arbitrarily large squeezing parameter $s$. Note that when $s=0$, this protocol reduces to the coherent-state protocol.
\item $V_0$ sends $|\psi\rangle$ to $P$, and $V_1$ sends $\theta$ to $P$. $V_0$ and $V_1$ coordinate their transmission times to make sure that $|\psi\rangle$ and $\theta$ arrive at $P$ at the same time.
\item When $P$ receives $|\psi\rangle$ and $\theta$, he first applies $R(-\theta)$, then he performs a homodyne detection to measure the quadrature $q_X$, and immediately broadcasts the measurement result $\alpha'$ to $V_0$ and $V_1$.
\item If the verifiers receive $\alpha'$ at a time inconsistent with the position of $P$, the protocol fails and is terminated.
\item Through an authenticated classical channel, $V_0$ and $V_1$ compare the reported measurement results received by them. If they receive different results, the protocol fails and is terminated.
\item $V_0$, $V_1$, and $P$ repeat the above steps many times. They estimate the conditional variance
\be\label{eq19} \Delta = \left\langle \left( \alpha-\dfrac{\alpha'}{\sqrt{\eta}} \right)^2 \right\rangle \ee
where $\eta$ is the overall transmittance of the quantum channel.

If $\Delta$ is consistent with the value expected from a channel with an overall transmittance $\eta$, the location of $P$ is accepted as authenticated.
\end{enumerate}
In normal conditions (no adversaries), the expected conditional variance from the honest $P$ is given by
\be\label{eq20} \Delta_P = \dfrac{1}{4} e^{-2s} +\dfrac{1-\eta}{4\eta} \ee
where the second term on the right-hand side represents the vacuum noise due to the channel loss (referred to the input of the channel).

Again, the adversaries launch the attack described above by using a 50:50 beam splitter and homodyne detectors. Let us define the annihilation operators of the two input modes of the beam spliter as $a$ and $b$, and the annihilation operators of the two output modes as $c$ and $d$, respectively. They are related by $c=\dfrac{1}{\sqrt{2}}(a-b)$, and $d=\dfrac{1}{\sqrt{2}}(a+b)$.

$E_0$ sends $|\psi\rangle$ to mode $a$, the vacuum state to mode $b$, and measures $q_X$ of mode $c$ and $q_P$ of mode $d$,
\bes q_X^{(c)} &=& \dfrac{c^{+}+c}{2}=\dfrac{1}{2\sqrt{2}}(a^{+}+a-b^{+}-b) \nonumber\\
q_P^{(d)} &=& \dfrac{i(d^{+}-d)}{2}=\dfrac{i}{2\sqrt{2}}(a^{+}-a+b^{+}-b) \ees
After receiving information on $\theta$ from $V_1$, the adversaries report
$\alpha'=\sqrt{2\eta}(q_X^{(c)} \cos\theta + q_P^{(d)} \sin\theta)$.

Using Eq.\ \eqref{eq19} and the following relations
\bes R^\dagger (\theta) a R(\theta) &=& e^{i\theta} a\, \nonumber\\  D^\dagger(\alpha) a D(\alpha) &=& a + \alpha \ , \nonumber \\ S^\dagger(s) a S(s) &=& a \cosh s - a^\dagger \sinh s \ , \ees
it is straightforward to calculate the conditional variance due to the attack,
\be\label{eq23} \Delta_E = \dfrac{1}{4} e^{-2s} +\dfrac{1}{4} \ee
Comparing Eqs.\ \eqref{eq20} and \eqref{eq23}, it is easy to see that as long as the overall transmittance $\eta<0.5$, we will have $\Delta_E < \Delta_P$, which means the adversaries can reproduce the noise variance expected from the honest $P$ thus the PBQC protocol is not secure. It should be noted that this result is independent of the squeezing parameter $s$.

\section{Discussion}
\label{sec:dis}

In principle, both the BB84 QKD and the GMCS QKD can tolerate arbitrarily high channel loss when implemented with perfect devices. This is because Alice and Bob share an authenticated classical channel, which allows them to perform either post-selection or reverse reconciliation. However, in PBQC, the verifers and the prover do not share an authenticated classical channel, so they cannot apply the post-selection or reverse reconciliation scheme. This could explain why PBQC protocols based on the BB84 or GMCS encoding schemes cannot go beyond the $3$ dB loss limit.

A natural way to go beyond the above $3$ dB limit is to use multiple encoding bases. Our discussion showed that by introducing the decoy state idea, an extended BB84-type PBQC protocol implemented with a weak coherent source and realistic SPDs could tolerate a total loss of $13$ dB when the intrinsic QBER is $1\%$. Such a protocol could find real-life applications. To further extend its working distance, one could explore new encoding schemes using a large set of mutually unbiased bases such as the one employed in \cite{Toshihiko14}.

In the case of CV PBQC protocols based on coherent detection, we have been unable to find a good way to go beyond the $3$ dB limit, even by using arbitrarily squeezed states and perfect homodyne detectors. This result could be understood from the point of view of a virtual entanglement model \cite{Grosshans032}. In the squeezed state PBQC protocol discussed in Section \ref{sec:CV}, $V_0$ could prepare the quantum state $|\psi\rangle$ from a two-mode squeezed vacuum (TMSV) by measuring a randomly chosen quadrature $q_\theta$ of mode $I$ using a homodyne detector, and sends mode $II$ to $P$. Later on, $E_0$ measures both the $X$-quadrature and $P$-quadrature of mode $II$ by performing a conjugate homodyne detection. Since measurements performed by $V_0$ and $E_0$ commute with each other, we can reverse the order without changing the measurement statistics. In this picture, $E_0$ measures mode $II$ first while projecting mode $I$ onto a Gaussian-modulated coherent state. Then $V_0$ measures its quadrature $q_\theta$. As such, the squeezed-state protocol is equivalent to a coherent state protocol and suffers from the same limitations. Nevertheless, the limitation of CV PBQC protocol can be partially compensated by the relatively high efficiency of a practical homodyne detector. 

\acknowledgments{
We would like to thank Ryan Bennink, Hoi-Kwong Lo, and Pavel Lougovski for very helpful discussions. This work was performed at Oak Ridge National Laboratory, operated by UT-Battelle for the U.S. Department of Energy under Contract No. DE-AC05-00OR22725. B.Q. acknowledges support from the laboratory directed research and development program and the U.S. Department of Energy Cybersecurity for Energy Delivery Systems (CEDS) program.}

\appendix*

\begin{figure}[t]
	\includegraphics[width=.45\textwidth]{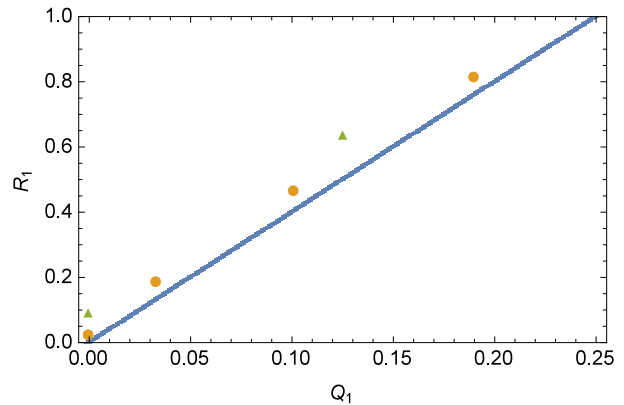}
	\caption{(Color online) Reporting rate ($R_1$) \emph{vs.}\ QBER ($Q_1$) for finite number of bases with $N=4$ (triangles), $N=8$ (circles), and $N\to\infty$ (solid line). } 
	\label{fig:1}
\end{figure}

\section{Finite number of bases}

The PBQC with finite bases is summarized below:
\begin{enumerate}
\item The verifiers agree on a random bit $x \in \{ 0,1 \}$ and two random numbers $\theta= \theta_m$ (where $m=0,1,\dots, N-1$), and $\phi$. The latter is chosen as follows. Given $m$ (i.e., for a fixed $\theta = \theta_m$), an integer $n\in \{0,1,\dots , [2N\sin \theta_m] \}$ is chosen randomly. Then the verifiers set $\phi = \phi_{m,n}$. The chosen values of the two angles are, respectively,
\be\label{A1} \theta_m = \frac{m\pi}{N} \ , \ \phi_{m,n} = \frac{\pi n}{N\sin\theta_m}  \ee
For large $N$, this is a uniform distribution over the Bloch sphere dividing its area into
\be K= \left[ N \left( 1+2 \cot \frac{\pi}{2N} \right) \right] \approx \left[ \frac{4N^2}{\pi} \right] \ee
square pixels each of side $\frac{\pi}{N}$.

$V_0$ prepares a qubit in the state
\be |\psi \rangle = U(\theta,\phi) |x\rangle \ee
where
$ U(\theta,\phi) = \left( \begin{array}{cc} \cos\frac{\theta}{2} & e^{i\phi}\sin\frac{\theta}{2} \\
 -e^{-i\phi}\sin\frac{\theta}{2} & \cos\frac{\theta}{2} \end{array} \right) $.
It should be noted that for $\theta =0$, all choices of $\phi$ correspond to the same basis, $\{ |0\rangle , |1\rangle \}$.
\item $V_0$ sends $|\psi\rangle$ to $P$, and $V_1$ sends $\theta$ and $\phi$ to $P$. $V_0$ and $V_1$ coordinate their transmission times to make sure that $|\psi\rangle$, $\theta$ and $\phi$ arrive at $P$ at the same time.
\item As soon as $|\psi\rangle$, $\theta$ and $\phi$ arrive, $P$ performs a measurement in the basis $\{ U(\theta,\phi) |0\rangle , U(\theta,\phi) |1\rangle \}$. If $P$ detects a photon successfully, he broadcasts the measurement result $x'$ back to $V_0$ and $V_1$ immediately. Otherwise, he reports no detection.
\item If the verifiers receive $x'$ at a time inconsistent with the position of $P$, the protocol fails and is terminated.
\item Through an authenticated classical channel, $V_0$ and $V_1$ compare the reported measurement results received by them. If they receive different results, the protocol fails and is terminated.
\item $V_0$, $V_1$ and $P$ repeat the above steps many times. They estimate the reporting rate and the QBER. For a given reporting rate, if the QBER is below a certain predetermined value, the location of $P$ is accepted as authenticated.
\end{enumerate}

We assume that $V_0$ employs a perfect single photon source to implement the above PBQC protocol. $E_0$ intercepts $|\psi\rangle$ and immediately performs a measurement $\{ \Pi_0, \mathbb{I} -\Pi_0 \}$, where $\Pi_0$ projects onto the state
\be |\theta_E, \phi_E\rangle = \cos\frac{\theta_E}{2} |0\rangle + e^{i\phi_E} \sin\frac{\theta_E}{2} |1\rangle \ee
She then forwards the measurement outcome $x_E$ to $E_1$. At the same time, $E_1$ intercepts $V_0$'s basis information $(\theta,\phi)$ from $V_1$ and forwards it to $E_0$. Then the adversaries have three options (agreed upon from the outset):
\begin{itemize}
\item[{\em (a)}] Report $x_E$ ($E_0$ to $V_0$ and $E_1$ to $V_1$), if $(\theta,\phi) \in S_a$.
\item[{\em (b)}] Report $(x_E+1)\text{mod}2$, if $(\theta,\phi) \in S_b$.
\item[{\em (c)}] Report no detection, if $(\theta,\phi) \in S_c$.
\end{itemize}
where $S_a$, $S_b$, $S_c$ are disjoint sets forming a partition of the set of possible pairs $\{\theta_m,\phi_{m,n}\}$ (or equivalently, the pixels on the Bloch sphere). 

The reporting rate is
\be R_1 = \frac{\# (S_a \bigcup S_b)}{K} = 1 - \frac{\# (S_c)}{K}\ee
The probability of error is
\be\label{A5} p(\theta,\phi)= \left\{ \begin{array}{ccc}  \sin^2 \frac{\theta - \theta_E}{2} + \sin^2 \frac{\phi - \phi_E}{2} \sin\theta \sin\theta_E & , & (\theta,\phi) \in S_a\\
\cos^2 \frac{\theta + \theta_E}{2} + \cos^2 \frac{\phi - \phi_E}{2} \sin\theta \sin\theta_E & , & (\theta,\phi) \in S_b \\
0 & , & (\theta,\phi) \in S_c\end{array} \right. \ee

The average QBER in the adversaries' attack is
\be Q_1 = \frac{1}{R_1 K} \sum_{n,m} p_{m,n} \ \ , \ \ \ \ p_{m,n} \equiv p(\theta_m, \phi_{m,n})\ee
In the above equations, the subscript $1$ indicates that a single photon source is applied.

Since the distribution given in \eqref{A1} is uniform, the adversaries' choice of angles $(\theta_E, \phi_E)$ does not affect the performance of their strategy. They need to choose them randomly, in order for their attack not to be detected. However, for our calculations, we may choose $\theta_E = 0$, $\phi_E =0$, without loss of generality. In this case, the probability of error becomes independent of $\phi$. It is given by
\be p_{m,n} = \left\{ \begin{array}{ccc}  \sin^2\frac{\theta_m}{2} & , & (\theta_m,\phi_{m,n} ) \in S_a\\
\cos^2\frac{\theta_m}{2} & , & (\theta_m,\phi_{m,n}) \in S_b \\
0 & , & (\theta_m,\phi_{m,n}) \in S_c\end{array} \right. \ee
Notice that $sin^2 \frac{\theta_m}{2} < \cos^2 \frac{\theta_m}{2}$ for $\theta_m < \frac{\pi}{2}$, whereas the opposite holds for $\theta_m > \frac{\pi}{2}$. Moreover, $\sin^2 \frac{\theta}{2}$ is monotonically increasing in the interval $[0,\frac{\pi}{2} )$. It follows that the best reporting strategy of the adversaries is to report all outcomes for $m \le m_0$, and opposite outcomes for $m \ge N-m_0$ where $m_0 \le N/2$. This leads to the optimal reporting rate
\bes R_1 &=& \frac{2}{K} \sum_{m=0}^{m_0} [1+2N\sin \theta_m ] \nonumber\\
&\approx&  \frac{2(m_0+1)}{K} + 2 \sin \frac{m_0 \pi}{2N} \sin \frac{(m_0+1)\pi}{2N} \ees
The minimum reporting rate is $R_1 = \frac{2}{K}$, as expected (found by setting $m_0 =0$), whereas the maximum reporting rate $R_1 =1$ is obtained by setting $m_0 \approx \frac{N}{2}$.

The corresponding QBER in this optimal strategy for the adversaries is
\bes Q_1 &=& \frac{2}{R_1K} \sum_{m=0}^{m_0} [ 1+ 2N\sin \theta_m ] \sin^2 \frac{\theta_m}{2} \nonumber\\
&\approx& \frac{\frac{2m_0+1}{4}+ \frac{2N^2}{\pi} \sin^2 \frac{m_0\pi}{2N} \sin^2 \frac{(m_0+1)\pi}{2N} -\frac{N}{2\pi} \sin \frac{(2m_0+1)\pi}{2N}}{ m_0+1+ \frac{4N^2}{\pi} \sin \frac{m_0\pi}{2N} \sin \frac{(m_0+1)\pi}{2N}  } \nonumber\\
\ees
At the minimum reporting rate $R_1=\frac{2}{K}$, we obtain $Q_1 =0$, as expected. At the maximum reporting rate, $R_1=1$, we obtain $Q_1 \approx \frac{1}{4}$. As $N\to\infty$, we recover our earlier results in the case of continuous distribution of bases. We obtain $R_1 \approx 2\sin^2 \frac{m_0\pi}{2N}$, which is in agreement with eq.\ \eqref{eq3} with $\Theta_0 = \frac{m_0\pi}{N}$. Morever, after some algebra, we obtain asymptotically, $Q_1 \approx \frac{1}{4} R_1$, in agreement with eq.\ \eqref{eq4}. Results for $N=4,8$ and $N\to\infty$ are shown in figure \ref{fig:1}. The convergence to the limit of a continuous distribution is fast -- even the case $N=8$ is practically indistinguishable from the limit $N\to\infty$.

\end{document}